\documentclass[11pt,letterpaper]{article}

\newcommand\nc{\newcommand}

\nc{\textsize}[2]{ \textwidth=#1 \oddsidemargin=8.5in
\addtolength\oddsidemargin{-2in}
\addtolength\oddsidemargin{-\textwidth} \divide\oddsidemargin by2
\evensidemargin=\oddsidemargin \textheight=#2 \topmargin=11in
\addtolength\topmargin{-2in} \addtolength\topmargin{-\textheight}
\divide\topmargin by2 }

\newtheorem{theorem}{Theorem}[section]
\newtheorem{lemma}[theorem]{Lemma}
\newtheorem{cor}[theorem]{Corollary}
\newtheorem{fact}[theorem]{Fact}

\nc{\crl}[2]{\begin{cor}\label{crl:#1} #2 \end{cor}}
\nc{\lem}[2]{\begin{lemma}\label{lem:#1} #2 \end{lemma}}
\nc{\thm}[2]{\begin{theorem}\label{thm:#1} #2 \end{theorem}}
\nc{\fct}[2]{\begin{fact}\label{fct:#1} #2 \end{fact}}

\nc{\refc}[1]{Corollary~\ref{crl:#1}}
\nc{\refl}[1]{Lemma~\ref{lem:#1}}
\nc{\reft}[1]{Theorem~\ref{thm:#1}}
\nc{\reffct}[1]{Fact~\ref{fct:#1}}

\def\qed{\rule{1.5mm}{3mm}}
\nc{\proof}[1]{ {\bf Proof.} #1 \hfill \qed\par}

\textsize{5.3in}{8.8in}

\title{\large \bf Tight Approximation Ratio of a General Greedy Splitting
Algorithm for the Minimum $k$-Way Cut Problem}

\author{{\sc Mingyu Xiao} ~~~~~  {\sc Leizhen Cai} ~~~~~ {\sc Andrew C. Yao} \\ \\
\normalsize Department of Computer Science and Engineering\\
\normalsize The Chinese University of Hong Kong\\
\normalsize Hong Kong SAR, CHINA\\ \\
\small Email: {\tt myxiao(lcai)@cse.cuhk.edu.hk, andrewcyao@tsinghua.edu.cn} \\
}
\date{}

\begin{document}
\maketitle

\begin{abstract}
For an edge-weighted connected undirected graph, the minimum
$k$-way cut problem is to find a subset of edges of minimum total
weight whose removal separates the graph into $k$ connected
components. The problem is NP-hard when $k$ is part of the input
and W[1]-hard when $k$ is taken as a parameter.

A simple algorithm for approximating a minimum $k$-way cut is to
iteratively increase the number of components of the graph by
$h-1$, where $2 \le h \le k$, until the graph has $k$ components.
The approximation ratio of this algorithm is known for $h \le 3$
but is open for $h \ge 4$.

In this paper, we consider a general algorithm that iteratively
increases the number of components of the graph by $h_i-1$, where
$h_1 \le h_2 \le \cdots \le h_q$ and $\sum_{i=1}^q (h_i-1) = k-1$.
We prove that the approximation ratio of this general algorithm is
$2 - (\sum_{i=1}^q {h_i \choose 2})/{k \choose 2}$, which is
tight. Our result implies that the approximation ratio of the
simple algorithm is $2-h/k + O(h^2/k^2)$ in general and $2-h/k$ if
$k-1$ is a multiple of $h-1$.

\vspace*{5mm} \noindent {\bf Key words} \ approximation algorithm,
$k$-way cut, $k$-way split.
\end{abstract}

\newpage

\section{Introduction}

Let $G =(V,E;w)$ a connected undirected graph with $n$ vertices
and $m$ edges, where each edge $e$ has a positive weight $w(e)$,
and $k$ a positive integer. A \emph{$k$-way cut} of $G$ is a
subset of edges whose removal separates the graph into $k$
connected components, and the \emph{minimum $k$-way cut} problem
is to find a $k$-way cut of minimum total weight. We note that
$k$-way cuts are also referred to as \emph{$k$-cuts} or
\emph{multi-component cuts} in the literature.

The minimum $k$-way cut problem is a natural generalization of the
classical \emph{minimum cut} problem and has been very well
studied in the literature. Goldschmidt and
Hochbaum~\cite{polynominalKCut} proved that the minimum $k$-way
cut problem is NP-hard when $k$ is part of the input and gave an
$O(n^{(1/2-o(1))k^2})$ algorithm. Kamidoi et
al.~\cite{kamidoi:kcut} presented an $O(n^{4k/(1 - 1.71/\sqrt k )
- 31})$ algorithm, and Xiao~\cite{xiao:kway} presented an
$O(n^{4k-\log k})$ algorithm. These three algorithms are based on
a divide-and-conquer method. Karger and
Stein~\cite{KargerStein:MinimumCut} proposed a randomized
algorithm that runs in $O(n^{2k-2}\log^3n)$ expected time.
Recently, Thorup~\cite{thorup:kway} obtained an $O(n^{2k}\log n)$
algorithm via tree packing. On the other hand, Downey et
al.~\cite{Downey:CuttingUpisHard} showed that the problem is
W[1]-hard when $k$ is taken as a parameter, which indicates that
it is very unlikely to solve the problem in $f(k)n^{O(1)}$ time
for any function $f(k)$. We also note that faster algorithms are
available for small $k$. Nagamochi and
Ibaraki~\cite{NagamochiIbaraki}, and Hao and
Orlin~\cite{Hao:mincut} solved the minimum 2-way cut problem
(i.e., the minimum cut problem) in $O(mn + n^2\log n)$ and
$O(mn\log(n^2/m))$ time respectively. Burlet and
Goldschmidt~\cite{Burlet:3cut} solved the minimum 3-way cut
problem in $\widetilde O(mn^3)$ time, Nagamochi and
Ibaraki~\cite{NagamochiIbaraki3-4-waycut} gave $\widetilde
O(mn^k)$ algorithms for $k\leq4$, and Nagamochi et
al.~\cite{nagamochi99:5-wayand6-way} extended this result for
$k\leq 6$. Furthermore, Levine~\cite{Levin:3456waycut} obtained
$O(mn^{k-2}\log^3n)$ randomized algorithms for $k\leq 6$.

In terms of approximation algorithms, Saran and
Vazirani~\cite{kcut2approximation} gave two simple algorithms of
approximation ratio $2-2/k$. Naor and Rabani~\cite{Naor:kcut}
obtained an integer program formulation of this problem with
integrality gap 2, and Ravi and
Sinha~\cite{Ravi:kcutnetworkstrength} also derived a
2-approximation algorithm via the network strength method.

A simple algorithm~\cite{kcut2approximation} for approximating a
minimum $k$-way cut is to iteratively increase the number of
components of the graph by $h-1$, where $2 \le h \le k$, until the
graph has $k$ components. This algorithm has an approximation
ratio of $2-2/k$ for $h = 2$~\cite{kcut2approximation}, and
Kapoor~\cite{Kapoor:3-cuts} claimed that it achieves ratio
$2-\alpha(h,k)$ for $h\geq3$, where
$\alpha(h,k)=h/k-(h-2)/k^2+O(h/k^3)$. Unfortunately, his proof for
$h\geq3$ is incomplete. Later, Zhao et
al.~\cite{Zhao:approximatingk-way} established Kapoor's claim for
$h=3$: the ratio is $2-3/k$ for odd $k$ and $2-(3k-4)/(k^2-k)$ for
even $k$. However, for $h\geq 4$, it seems quite difficult to
analyze the performance of this algorithm and it has been an open
problem whether we get a better approximation ratio with this
approach.

In this paper, we consider a general algorithm that iteratively
increases the number of components of the graph by $h_i-1$, where
$h_1 \le h_2 \le \cdots \le h_q$ and $\sum_{i=1}^q (h_i-1) = k-1$.
We prove that the approximation ratio of this general algorithm is
$2 - (\sum_{i=1}^q {h_i \choose 2})/{k \choose 2}$, which is
tight. Our result implies that the approximation ratio of the
simple algorithm is $2-h/k + O(h^2/k^2)$ in general and $2-h/k$ if
$k-1$ is a multiple of $h-1$, which settles the open problem
mentioned earlier in the affirmative.

The rest of the paper is organized as follows. In
Section~\ref{algorithm}, we formalize our general greedy splitting
algorithms and present our main results on their approximation
ratios. We prove our main results in Section~\ref{analysis} while
the proof of a purely analytical lemma is given in
Section~\ref{addlemma}, and conclude with some remarks in
Section~\ref{remark}.

\section{Algorithms and main results}
\label{algorithm}

\vspace*{-1mm}In this section, we formalize our greedy splitting
algorithms and present our main results on their approximation
ratios. We note that Zhao et
al.~\cite{zhao:greedysplitting1,zhao:greedysplitting2} have
studied such algorithms for general multiway cut and partition
problems. First we extend the notion of $k$-way cuts to
disconnected graphs. A {\em $k$-way split} of a graph is a subset
of edges whose removal increases the number of components by
$k-1$. Therefore for a connected graph, a $k$-way split is
equivalent to a $k$-way cut. We note that the time for finding a
minimum $k$-way split in a general graph is the same as finding a
$k$-way cut~\cite{Zhao:approximatingk-way}.

One general approach for finding a light $k$-way cut is to find
minimum $h_i$-way splits successively for a given sequence
$(h_1,h_2,\cdots,h_q)$.

\noindent \textbf{Algorithm
iterative-split$(G,k,(h_1,h_2,\cdots,h_q))$}

\noindent Input: Connected graph $G = (V,E;w)$, integer $k$ and
sequence $(h_1,h_2,\cdots,h_q)$ of integers satisfying $2\leq
h_1\leq h_2 \leq \cdots \leq h_q$ and $\sum_{i=1}^q (h_i-1) =
k-1$.

\noindent Output: A $k$-way cut of $G$.

\vspace{-0.4cm}

\begin{enumerate}
\item   For $i := 1$ to $q$ find a minimum $h_i$-way split $C_i$
of $G$ and let $G\leftarrow G-C_i$.

\item  Return $\bigcup_{i=0}^qC_i$ as a $k$-way cut.
\end{enumerate}

\vspace*{-4mm}A special case of the above algorithm is when all
$h_i$'s in the integer sequence, with the possible exception of
the first one, are equal. The following gives a precise
description of this special case.

\noindent \textbf{Algorithm iterative-$h$-split$(G,k,h)$}

\noindent Input: Connected graph $G = (V,E;w)$, integers $k$ and
$h$.

\noindent Output: A $k$-way cut of $G$.

\vspace{-0.4cm}
\begin{enumerate}
\item Let $p = \lfloor \frac{k-1}{h-1} \rfloor$ and $r = (k-1)
\mbox{ \rm mod } (h-1)$.

\item If $r\neq 0$, then find a minimum $(r+1)-$way split $C_0$ of
$G$ and let $G\leftarrow G-C_0$.

\item For $i := 1$ to $p$ find a minimum $h$-way split $C_i$ of
$G$ and let $G\leftarrow G-C_i$.

\item Return $\bigcup_{i=0}^pC_i$ as a $k$-way cut.
\end{enumerate}

The above two algorithms run in polynomial time if $h_q$ and $h$
are bounded by some constant, and our main results of the paper
are the following two tight bounds for their approximation ratios.

\thm{split}{The approximation ratio of algorithm {\bf
iterative-split} is
\[ 2 - \frac{\sum_{i=1}^q {h_i \choose 2}}{{k \choose 2}}. \]
}

\crl{h-split}{The approximation ratio of algorithm {\bf
iterative-$h$-split} is
\[ 2- {h \over k} + {(h-1-r)r \over k(k-1)} = 2-{h \over k} + O({h^2 \over k^2}), \]
where $r = (k-1) \mbox{ \rm mod } (h-1)$. }

\noindent {\bf Remark.}{ We note that when $k-1$ is a multiple of
$h-1$, {\bf iterative-$h$-split} is a $(2-h/k)$-approximation
algorithm, and \refc{h-split} for $h=3$ yields a result of Zhao et
al.~\cite{Zhao:approximatingk-way}. }

\section{Performance analysis}
\label{analysis}

In this section, we will prove our main results on the
approximation ratios of our approximation algorithms. For this
purpose, we first establish a relation between the weight $w(C_h)$
of a minimum $h$-way split $C_h$ and the weight $w(C_k)$ of a
$k$-way split $C_k$, which will be the main tool in our analysis.
For convenience, we allow $h = 1$ (note that a minimum 1-way split
is an empty set). For a collection of mutually disjoint subsets
$V_1,V_2,\cdots,V_t \in V$, we use $[V_1,V_2,\cdots,V_t]$ to
denote the set of edges $uv$ such that $u \in V_i$ and $v \in V_j$
for some $V_i \not= V_j$.

\lem{two-cuts}{Let $G$ be an edge-weighted graph, $h \ge 1$, and
$k \ge \max\{2,h\}$. For any minimum $h$-way split $C_h$ and any
$k$-way split $C_k$ of $G$, the following holds.
\begin{eqnarray}\label{eq-1}
{w(C_h) \over w(C_k)} \leq (2 - {h \over k}){h-1 \over k-1}.
\end{eqnarray}
} \proof{First we consider the case that $G$ is connected. In this
case, $C_k$ and $C_h$, respectively, are $k$-way and minimum
$h$-way cuts of $G$, and thus $C_k$ corresponds to a partition
$\Pi = \{V_1,V_2,\dots, V_k\}$ of the vertex set $V$ of $G$ such
that each $V_i$ is a component of $G-C_k$.

We can merge any $k - (h -1)$ elements in $\Pi$ into one element
to form a new partition $\Pi' = \{V'_1,V'_2, \dots, V'_h\}$ of
$V$. Let $E(\Pi') = [V'_1,V'_2, \dots, V'_h]$. Then $G - E(\Pi')$
has at least $h$ components, and therefore the weight $w(E(\Pi'))$
of $E(\Pi')$ is at least $w(C_h)$. There are ${k \choose h-1}$
different ways to form $\Pi'$, and therefore the total weight $W$
of all $E(\Pi')$ is at least ${k \choose h-1} w(C_h)$.

On the other hand, we can put an upper bound on $W$ by relating it
to the weight of $C_k$. Consider the set $E_{ij}$ of edges in
$C_k$ between $V_i$ and $V_j$. For a partition $\Pi'$, $E_{ij}
\subseteq E(\Pi')$ iff $V_i$ and $V_j$ are not merged in forming
$\Pi'$. The number of $\Pi'$s for which $V_i$ and $V_j$ are merged
is ${k-2 \choose h-1}$, implying that each $E_{ij}$ is counted ${k
\choose h-1} -{k-2 \choose h-1}$ times in calculating $W$.
Therefore
\[ W  = ({k \choose h-1} -{k-2 \choose h-1}) \cdot w(C_k)
\ge {k \choose h-1} \cdot w(C_h), \] which yields the inequality
in the lemma.

For the case that $G$ is disconnected, we construct a connected
graph $G' = (V',E'; w')$ from $G$ as follows:
\begin{enumerate}
\item Add a new vertex $v$. \item For each component $H$ of $G$,
add an edge $e_H$ between $v$ and an arbitrary vertex of $H$.
\item Set the weight of $e_H$ to $\infty$. \item Set $w'(e) =
w(e)$ for all other edges of $G'$.
\end{enumerate}

Then every $k$-way split in $G$ is a $k$-way cut in $G'$, and
every minimum $h$-way split in $G$ is a minimum $h$-way cut in
$G'$. Since $G'$ is connected, the lemma holds for $G'$ and hence
for $k$-way and minimum $h$-way splits of $G$. }

For convenience, define for all $h \geq 1$ and $k \geq \max\{2,
h\}$,
\[ f(k,h) = (2 - {h \over k}){h-1 \over k-1}. \]
We note that the bound in \refl{two-cuts} is tight, which can be
seen by considering a $k$-way cut and a minimum $h$-way cut of the
complete graph $K_k$. This also gives a combinatorial explanation
of $f(k,h)$: the ratio between the number of edges covered by
$h-1$ vertices in $K_k$ and the number of edges of $K_k$. We also
need the following properties of $f(k,h)$ in our analysis.

\fct{func}{Function $f(k,h)$ monotonically increases for $h \in
[1, k]$ and monotonically decreases for $k \in [h,\infty)$.}

\fct{prod}{For all $a\geq0, h\geq 2$, and $k\geq a+h,$
\begin{eqnarray}
f(k-a,h)(1-f(k,a+1))\leq f(k,h).
\end{eqnarray}}
\proof{Straightforward manipulation gives
\begin{eqnarray*}
    f(k-a,h)(1-f(k,a+1)) = (2- {{2a+h} \over k}){h-1 \over k-1} \leq f(k,h).
\end{eqnarray*} }

The next inequality is an analytical result critical to the proof
of our main theorem. Let $q \ge 2$. For any integers $2 \leq h_1
\leq h_2 \leq \cdots \leq h_q$, $0 \leq a \leq h_1 -1$ and
$k-1\geq \sum_{i=1}^q (h_i-1)$, let
\begin{equation}
\label{D} D=f(k-a, h_1-a) + \sum_{i=2}^q f(k-a, h_i)
\end{equation} and
\begin{equation}
F= \max \{ D, \ \  f(k, a+1) + (1-f(k,a+1)) D \}.
\end{equation}

\lem{F}{$F \leq \sum_{i=1}^qf(k,h_i)$.}

To avoid distraction from our main discussions, we delay the proof
of this purely analytical lemma to Section~\ref{addlemma}.

We are now ready to prove our main results. For this purpose, we
call a sequence $((C_1,h_1),\dots,(C_q,h_q))$ a {\em nondecreasing
$q$-sequence of minimum splits} if integers $2 \le h_1 \le h_2 \le
\cdots \le h_q$ and each $C_i$, $1 \le i \le q$, is a minimum
$h_i$-way split of $G_{i}=G-\bigcup_{j=1}^{i-1} G_j$. To prove
\reft{split}, it suffices to prove the following theorem. We note
that although the proof is an inductive one, the argument in the
proof is subtle, and the condition $h_1 \leq h_2 \leq \cdots \leq
h_q$ is crucial to the proof.

\thm{ratio}{Let $((C_1,h_1),\dots,(C_q,h_q))$ be a nondecreasing
$q$-sequence of minimum splits of a weighted graph $G= (V,E;w)$,
where $w : E \rightarrow R^+$, and $S_k$ a $k$-way split of $G$
satisfying $k-1\geq \sum_{i=1}^q (h_i-1)$. Then
\begin{eqnarray}\label{eq-lem3}
w(\bigcup_{i=1}^q C_i) \leq \sum_{i = 1}^q f(k,h_i)\cdot w(S_k).
\end{eqnarray}
} \proof{We use induction on $q$. For $q=1$, the theorem is
established by \refl{two-cuts}. For the inductive step, let $q \ge
2$, $C'_1 = C_1 \cap S_k$, $S_{k'} = S_k - C'_1$, and $C''_1 = C_1
- C'_1$. Then $C'_1$ is an $(a+1)$-way split of $G$ for some $0
\le a \le h_1-1$, $C''_1$ is a minimum $(h_1-a)$-way split of
$G-C'_1$ (otherwise $C_1$ would not be a minimum $h_1$-way split
of $G$), and $S_{k'}$ is a $(k-a)$-way split of $G - C'_1$. It
follows that $S_{k'}$ is a $k'$-way split of $G - C_1$ for some
$k' \ge k -a$. Note that $((C_2,h_2),\dots,(C_q,h_q))$ is a
nondecreasing $(q-1)$-sequence of minimum splits of $G - C_1$ and
$k'-1\geq \sum_{i=2}^q (h_i-1)$. By the induction hypothesis and
the fact that each $f(k',h_i)$ is at most $f(k-a,h_i)$
(\reffct{func}), we have
\begin{eqnarray}\label{eq-hyp}
w(\bigcup_{i=2}^q C_i) \leq \sum_{i = 2}^q {f(k',h_i)\cdot
w(S_{k'})}
        \leq \sum_{i = 2}^q {f(k-a,h_i)\cdot w(S_{k'})}.
\end{eqnarray}
Let $W = w(C_1) + \sum_{i = 2}^q {f(k-a,h_i)\cdot w(S_{k'})}$.
Then $w(\bigcup_{i=1}^q C_i) \le W$ by (\ref{eq-hyp}), and we will
establish the theorem by proving $W \le \sum_{i = 1}^q
{f(k,h_i)\cdot w(S_k)}$.

If $w(C_1')> f(k, a+1) w(S_k)$, then $w(S_{k'})=w(S_k)-w(C_1')
\leq (1-f(k, a+1))w(S_k)$. By \refl{two-cuts}, we have
$w(C_{1})\leq f(k,h_1)\cdot w(S_k)$ and it follows from
\reffct{prod} that
\begin{eqnarray*}
W & \le & (f(k,h_1)+\sum_{i =2}^q {f(k-a,h_i)(1-f(k,a+1))}) \cdot w(S_k)  \\
  & \le & \sum_{i = 1}^q {f(k,h_i)\cdot w(S_k)}.
\end{eqnarray*}

Otherwise, $w(C'_1) \leq f(k,a+1)\cdot w(S_k)$ and we have
\[ W  =  w(C'_1) + w(C''_1) + \sum_{i = 2}^q f(k-a,h_i)\cdot w(S_{k'}). \]
Since $C''_1$ is a minimum $(h_1-a)$-way split of $G-C'_1$, we
have $w(C''_{1})\leq f(k-a, h_1-a)\cdot w(S_{k'})$ by
\refl{two-cuts}. It follows that
\begin{eqnarray*}
W & \le & w(C'_1) + f(k-a, h_1-a) \cdot w(S_{k'}) + \sum_{i = 2}^q f(k-a,h_i)\cdot w(S_{k'})\\
 & = & w(C'_1) + D \cdot w(S_{k'})
\end{eqnarray*}
for $D= f(k-a, h_1-a) + \sum_{i = 2}^q f(k-a,h_i)$ as defined in
(\ref{D}). Define $x=w(C_1')/w(S_k)$ and we have $W \le (x+
(1-x)D)w(S_k)$. Since $0 \leq x \leq f(k, a+1)$, the maximum value
of $x+(1-x)D$ over the interval $[0,f(k, a+1)]$ must be at either
$x=0$ or $x=f(k,a+1)$ as it is a linear function in $x$. This
means
\[ {W \over w(S_k)} \leq  \max \{ D, f(k, a+1) + (1-f(k,a+1)) D\}.\]
Therefore by \refl{F}, we have
\[ W  \leq (\sum_{i=1}^qf(k,h_i))\cdot w(S_k).\]
This completes the inductive step and therefore proves the
theorem. }

We can obtain \reft{split} for Algorithm {\bf iterative-split}
from \reft{ratio} as follows (note that $\sum_{i = 1}^q (h_i-1) =
k-1$):
\begin{eqnarray*}
\sum_{i = 1}^q f(k,h_i) & = & \sum_{i = 1}^q (2- \frac{h_i}{k})\frac{h_i-1}{k-1} \\
    & = & \frac{2}{k-1} \sum_{i = 1}^q (h_i-1)
        - \frac{1}{k(k-1)} \sum_{i = 1}^q h_i(h_i-1)  \\
    & = & 2 - \frac{\sum_{i = 1}^q {h_i \choose 2}}{{k \choose 2}}. \\
\end{eqnarray*}
For Algorithm {\bf iterative-$h$-split}, we can easily derive
\refc{h-split} from \reft{split}.

\noindent {\bf Remark} The bound in \reft{ratio} is tight for $k-1
= \sum_{i=1}^q (h_i-1)$ and therefore the approximation ratios in
\reft{split} and \refc{h-split} are tight. To see this, consider
the following graph $G$ that consists of the disjoint union of
$q+1$ copies $H_1, H_2, \cdots, H_q, K$ of the complete graph
$K_k$. For each $H_i$, fix a subset $V_i$ of $h_i-1$ vertices and
let $E_i$ denote edges in $H_i$ that are covered by $V_i$. Each
edge in $E_i$ has weight 1, and each of the remaining edges of
$H_i$ has weight $\infty$. Set the weight of every edge in $K$ to
1.

A minimum $k$-way split $C_k$ of $G$ consists of all edges in $K$,
but {\bf iterative-split} may return $\bigcup_{i=1}^q E_i$ as a
$k$-way split $C'_k$ of $G$. Since $w(C_k) = {k \choose 2}$ and
$w(C'_k) = \sum_{i=1}^q |E_i| = f(k,h_i){k \choose 2}$, we have
$w(C'_k)/w(C_k) = \sum_{i=1}^q f(k,h_i)$.

\section{Proof of \refl{F}}
\label{addlemma}

In this section, we complete our performance analysis by proving
\refl{F}: $F \leq \sum_{i=1}^qf(k,h_i)$, where $F= \max \{ D, \ \
W'\}$ for $ D=f(k-a, h_1-a) + \sum_{i=2}^q f(k-a, h_i)$ and
$W'=f(k, a+1) + (1-f(k,a+1)) D$. For this purpose, we first derive
some useful properties of $f(k,h)$.

\fct{two-sets}{For all $h_1, h_2\geq 0$ and $k\geq
\max\{h_1+h_2+1,2\}$,
\[ f(k,h_1+h_2+1)= f(k,h_1+1)+f(k-h_1,h_2+1)(1-f(k,h_1+1)). \]
} \proof{Let $e(k,h)$ denote the number of edges covered by $h$
vertices in the complete graph $K_k$, and $m_k$ the number of
edges in $K_k$. Then
\[ e(k, h_1 + h_2) = e(k, h_1) + e(k-h_1, h_2),\]
and thus
\[ \frac{e(k, h_1 + h_2)}{m_k}  =  \frac{e(k, h_1)}{m_k} +
    \frac{e(k-h_1, h_2)}{m_{k-h_1}} \cdot \frac{m_{k-h_1}}{m_k}. \]
Since $m_{k-h_1} = m_k - e(k, h_1)$, we obtain
\[ \frac{e(k, h_1 + h_2)}{m_k}  =  \frac{e(k, h_1)}{m_k} +
        \frac{e(k-h_1, h_2)}{m_{k-h_1}} \cdot (1 - \frac{e(k, h_1)}{m_k}), \]
and the lemma follows from the fact that $f(k,h) = e(k,h-1)/m_k$.
}

\fct{diff}{ For all $a\geq0, h_2\geq h_1 \ge 2, k\geq a+h_2,$
\[ f(k-a,h_2)-f(k,h_2)\leq {\frac{{h_2-1}}{h_1-1}}[f(k-a,h_1)-f(k,h_1)]. \]
} \proof{
\begin{eqnarray*}
   & \Leftrightarrow  & {f(k-a,h_2)-{\frac{{h_2-1}}{h_1-1}}f(k-a,h_1)\leq f(k,h_2)-{\frac{{h_2-1}}{h_1-1}}f(k,h_1)}  \\
   & \Leftrightarrow  & {{-\frac{(h_2-h_1)(h_2-1)}{(k-a)(k-a-1)}}\leq {-\frac{(h_2-h_1)(h_2-1)}{k(k-1)}}  }  \\
   & \Leftrightarrow  & {(k-a)(k-a-1)\leq k(k-1)}.
\end{eqnarray*}  }

\fct{sum}{ For all $a\geq0, h\geq2, k\geq a+h,$
\[ f(k-a,h-a)+{\frac{k-h}{h-1}}f(k-a,h)\leq {\frac{{k-1}}{h-1}}f(k,h). \]
} \proof{
\begin{eqnarray*}
    &\Leftrightarrow  & {\frac{a^2 + a(1+2h-4k) - (h - 2k)(k-1)}{(k-a)(k-a-1)}}\leq \frac{2k-h}{k}  \\
    &\Leftrightarrow  & k(a^2 + a(1+2h-4k) - (h - 2k)(k-1))\leq (2k-h)(k-a)(k-a-1)  \\
    &\Leftrightarrow  & a(a+1)(h-k)\leq 0.
\end{eqnarray*}  }

\fct{sumup}{For all $2 \le h_1\leq h_i$ $(i=2,3,\cdots,q)$, $0\leq
a< h_1$, $\sum_{i = 1}^q{(h_i-1)}\leq k-1 $,
\[ f(k-a,h_1-a)+\sum\limits_{i = 2}^q f(k-a,h_i)\leq
f(k,h_1)+\sum\limits_{i = 2}^q f(k,h_i). \] } \proof{Let
$\Delta=f(k-a,h_1-a)+\sum\limits_{i = 2}^q
f(k-a,h_i)-f(k,h_1)-\sum\limits_{i = 2}^q f(k,h_i)$. By
\reffct{diff}, we have
\begin{eqnarray*}
\sum\limits_{i = 2}^q (f(k-a,h_i)- f(k,h_i))
& \le & \sum\limits_{i = 2}^q \frac{h_i-1}{h_1-1} (f(k-a,h_1)-f(k,h_1)) \\
& = & \frac{k-h_1}{h_1-1}(f(k-a,h_1)-f(k,h_1)). \\
\end{eqnarray*}
Therefore
\begin{eqnarray*}
\Delta &\leq&
f(k-a,h_1-a)-f(k,h_1)+\frac{k-h_1}{h_1-1}(f(k-a,h_1)-f(k,h_1))
\\
&=&
f(k-a,h_1-a)+\frac{k-h_1}{h_1-1}f(k-a,h_1)-\frac{k-1}{h_1-1}f(k,h_1).\\
\end{eqnarray*}
It follows from \reffct{sum} that $\Delta\leq 0$, which proves the
lemma. }

Now, we are ready to prove \refl{F}: $F \leq \sum_{i=1}^qf(k,
h_i)$. Recall that $F = \max\{D, W'\}$ for $D=f(k-a, h_1-a) +
\sum_{i=2}^q f(k-a, h_i)$ and $W'=f(k, a+1) + (1 - f(k, a+1))D$.
As $D \leq  \sum_{i=1}^qf(k, h_i)$ by \reffct{sumup}, we need only
show that $W'\leq \sum_{i=1}^qf(k,h_i)$. This can be done by using
\reffct{two-sets} and \reffct{prod} as follows:
\begin{eqnarray*}
W'   &  =    & f(k,a+1)-f(k-a,h_1-a)f(k,a+1)+f(k-a, h_1-a) \\
   &   & + \sum_{i =2}^q {f(k-a,h_i)}(1-f(k,a+1))\\
   & = & f(k,h_1)+\sum_{i =2}^q {f(k-a,h_i)}(1-f(k,a+1))
            \mbox{\rm \ \ \ (by \reffct{two-sets})} \\
   & \leq & \sum_{i = 1}^q {f(k,h_i)}.
            \mbox{\rm  \ \ \ (by \reffct{prod})} \\
\end{eqnarray*}

\section{Concluding remarks}
\label{remark}

In this paper, we have determined the exact approximation ratio of
a general splitting algorithm {\bf iterative-split} for the
minimum $k$-way cut problem. The answer is a surprisingly simple
expression $ 2-{\sum_{i=1}^q {h_i \choose 2}/{k \choose 2}}$, yet
it takes a somewhat subtle and involved inductive argument to
prove the result. It would be interesting to find a direct and
simpler proof.

We note that for {\bf iterative-split}, the requirement that $h_1
\le h_2 \le \cdots \le h_q$ is crucial for obtaining the
approximation ratio of the algorithm, which is unknown if we drop
the requirement. We also note that if we restrict $h_q$ to be at
most $h$, then {\bf iterative-$h$-split}, a special case of {\bf
iterative-split}, achieves the best approximation ratio among all
possible choices of $h_1 \le h_2 \le \cdots \le h_q$.

Finally, we may use {\bf iterative-split} as a general framework
for designing approximation algorithms for various cut and
partition problems, and the ideas in this paper may shed light on
the analysis of this general approach for these problems.

\end{document}